\documentclass[journal,twocolumn,10pt]{IEEEtran} 
\usepackage{epsfig,cite}
\usepackage{graphicx}
\usepackage{color}
\usepackage{amssymb,amsmath}
\definecolor{blue}{rgb}{0,0,1}
\definecolor{darkgreen}{rgb}{0,.5,0}
\definecolor{darkred}{rgb}{.5,0,0}

\def\given{\:|\:}

\newcommand{\lx}{\lambda_X}
\newcommand{\lw}{\lambda_W}

\begin{document}

\title{Entropy of the Sum of\\ Two Independent,
Non-Identically-Distributed\\ Exponential Random Variables}

%
\author{
	Andrew W. Eckford and Peter J. Thomas%
	\thanks{Andrew W. Eckford is with the Department of Electrical Engineering and Computer Science,
	York University, 4700 Keele Street, Toronto, Ontario, Canada M3J 1P3. Email: aeckford@yorku.ca}%
	\thanks{Peter J. Thomas is with the Department of Mathematics, Applied Mathematics, and Statistics, Department of Biology, and Department of Electrical and Computer Engineering, Case Western Reserve University, Cleveland, Ohio, USA. Email: pjthomas@case.edu}%
}

\maketitle

\begin{abstract}
In this letter, we give a concise, closed-form expression for the differential entropy of the 
sum of two independent, non-identically-distributed exponential random variables. The derivation
is straightforward, but such a concise entropy has not been previously given in the literature.
The usefulness of the expression is demonstrated with examples.
\end{abstract}

\section{Introduction}

Exponential distributions and Poisson point processes are widely used in communication and information theory: applications include queue-timing channels \cite{anantharam96, sundaresan06},
analysis of fading channels \cite{si14}, intercellular signal transduction \cite{eckford16},
and covert communication in networks \cite{cabruk04,wagner2005}.

For information-theoretic analysis,
the entropy of the exponential distribution, and the entropy of the
sum of two independent, identically distributed exponential random variables (which has the Erlang-2 distribution), are well known \cite{ver78}.
However,
we are particularly
motivated by Poisson processes in which the interval between arrivals is 
independent but {\em not
identically distributed:} since
the dwell time in one state of a Poisson process has the exponential distribution, the total
dwell time in two adjacent states is the sum of two non-identical exponential random variables. 

In equation (\ref{eqn:MainResult}), we give our main result, which is a concise, closed-form
expression for the entropy of the sum of two independent, non-identically-distributed exponential random variables. Beyond the specific applications given above, some of which are
illustrated with examples at the end of this letter, our result has broader applications to the calculation of entropy for continuous-time Markov chains or Poisson point processes \cite{mcfadden65}, and to the information theory of the exponential distribution in general
\cite{verdu96}. 

\section{Main Result}

We briefly review some known results about the exponential distribution, prior to stating
the main result in (\ref{eqn:MainResult}).

An exponentially-distributed random variable $X$ has probability density function (PDF)
\begin{equation}
	f_X(x) = \left\{ \begin{array}{cl} \lambda \exp(- \lambda x), & x \geq 0\\ 0, &x < 0, \end{array}\right. 
\end{equation}
where $\lambda > 0$ is a parameter, called the rate of the distribution.
The differential entropy of the exponential distribution is well known (we use the natural
logarithm throughout, so entropy is in nats):
\begin{align}
	h(X) &= - \int_0^\infty \lambda \exp(-\lambda x) \log \Big(\lambda \exp(-\lambda x) \Big) dx \\
	\label{eqn:SingleEntropy}
	&= 1 - \log \lambda .
\end{align}
This turns out to be the maximum entropy for a nonnegative random variable
with a mean constraint (see \cite[Ch. 12]{cover-book}).

Let $W$ and $X$ be independent and identically distributed (IID) exponential random variables
with rate $\lambda$. Further, let $Y = W+X$. Then $Y$ is known to have the Erlang-2 distribution. This distribution has differential entropy 
\cite{ver78}
\begin{equation}
	\label{eqn:IdenticalEntropy}
	h(Y) = 2 - \psi(2) - \log \lambda ,
\end{equation}
where $\psi$ is the digamma function (see \cite{weisstein-book}).

Now suppose $W$ and $X$ are independent and exponential, but {\em non-identically distributed}: 
the rates are $\lw$ and $\lx$, respectively. (Without loss of generality,
assume $\lw > \lx$.)
It is known that the distribution of $Y = W+X$ can be obtained from its characteristic function $\phi_Y(t)$ (with $j = \sqrt{-1}$):
\begin{align}
	\phi_Y(t) &= \phi_W(t) \phi_X(t) \\
	&= \frac{\lw}{\lw + jt} \frac{\lx}{\lx + jt} \\
	&= \frac{\lw \lx}{\lw - \lx}\left(\frac{1}{\lx + jt}
		- \frac{1}{\lw + jt}\right) ,
\end{align}
for which the corresponding PDF is
\begin{equation}
	\label{eqn:NonIdenticalPDF}
	f_Y(y) = \frac{\lw \lx}{\lw - \lx} \Big( \exp(-\lx y) - \exp(-\lw y) \Big).
\end{equation}
Our main result is to obtain the differential entropy $h(Y)$ of $f_Y(y)$ in (\ref{eqn:NonIdenticalPDF}).
We show that 
\begin{equation}
	\label{eqn:MainResult}
	h(Y) = 1 + \gamma + \log\left(\frac{\lw-\lx}{\lw\lx}\right) + \psi\left( \frac{\lw}{\lw - \lx} \right) 
\end{equation}
where $\gamma$ is the Euler gamma constant and $\psi$ is the digamma function.

The derivation of the result is straightforward, but the concise form of 
this entropy is not available in the literature, to the authors' knowledge.

\section{Derivation}


%
%
%

To simplify the expressions, let 
\begin{equation}
	c = \lw\lx / (\lw-\lx) .
\end{equation}
The differential entropy $h(Y)$ is obtained from the integral
\begin{align}
	\nonumber
	\lefteqn{h(Y)} &\\
	=&- \int_0^\infty c\left( e^{-\lambda_X y} - e^{-\lambda_W y} \right) \log c \left( e^{-\lambda_X y} - e^{-\lambda_W y} \right) dy \\
	\nonumber =& \: c\int_0^\infty e^{-\lambda_W y} \log c\left( e^{-\lambda_X y} - e^{-\lambda_W y} \right) dy\\ 
	& - c\int_0^\infty e^{-\lambda_X y} \log c\left( e^{-\lambda_X y} - e^{-\lambda_W y} \right) dy \\
	\nonumber =& \: c\int_0^\infty e^{-\lambda_W y} \log e^{-\lambda_X y}\left( 1 - e^{-(\lambda_W-\lambda_X )y} \right) dy \\ \nonumber
	& \:\:\:- c\int_0^\infty e^{-\lambda_X y} \log e^{-\lambda_X y}\left( 1 - e^{-(\lambda_W-\lambda_X )y} \right) dy \\
	& \:\:\:+ \frac{c\log c}{\lambda_W} - \frac{c\log c}{\lambda_X}\\
	\nonumber=&\:   c\int_0^\infty \lambda_X y e^{-\lambda_X y}dy 
	- c\int_0^\infty \lambda_X y e^{-\lambda_W y} dy\\
	\nonumber
	&	\:\:\:+ c\int_0^\infty e^{-\lambda_W y} \log \left( 1 - e^{-(\lambda_W-\lambda_X )y} \right) dy\\
	\nonumber&	\:\:\:- c\int_0^\infty e^{-\lambda_X y} \log \left( 1 - e^{-(\lambda_W-\lambda_X )y} \right) dy \\  \label{eqn:MainIntegral}&\:\:\:+ \frac{c\log c}{\lambda_W} - \frac{c\log c}{\lambda_X}.
\end{align}
From basic calculus, the first two terms in (\ref{eqn:MainIntegral}), on the first line, evaluate to
\begin{align}
	\nonumber & c\int_0^\infty \lambda_X y e^{-\lambda_X y}dy 
	- c\int_0^\infty \lambda_X y e^{-\lambda_W y} dy\\ 
	\label{eqn:FirstAndSecondTerms}
	& = \frac{c}{\lambda_X} - \frac{c\lambda_X}{\lambda_W^2} .
\end{align}

The middle two terms in (\ref{eqn:MainIntegral}), on the second and third lines, 
use an integral of the form
\begin{equation}
	\label{eqn:IntegralType}
	\int_0^\infty e^{-ux} \log (1 - e^{-vx}) dx .
\end{equation}
Performing a change of variables, let
\begin{equation}
	\xi = e^{-vx} ,
\end{equation}
which implies $x = -\frac{1}{v} \log \xi$ and $dx = -1/v \xi \:d\xi$.
Substituting into (\ref{eqn:IntegralType}),
\begin{align}
	\nonumber\lefteqn{\int_0^\infty e^{-ux} \log (1 - e^{-vx}) dx}&\\
	&= \int_1^0 \xi^{\frac{u}{v}} \log (1-\xi) \left(- \frac{1}{v\xi}\right) \:d\xi\\
	&= \frac{1}{v} \int_0^1 \xi^{\frac{u}{v} - 1} \log (1-\xi) \: d\xi .
\end{align}
A solution is given by Gradshteyn and Ryzhik \cite{gradshteyn-book}:
\begin{align}
	\nonumber\lefteqn{\frac{1}{v} \int_0^1 \xi^{\frac{u}{v} - 1} \log (1-\xi) \: d\xi} & \\
	&= \frac{1}{v} \left(- \frac{v}{u} \right) \left[ \psi\left(\frac{u}{v}+1\right) - \psi(1) \right] \\
	\label{eqn:GradRyz}
	&= - \frac{1}{u} \left[ \gamma + \psi\left(\frac{u}{v}+1\right) \right] ,
\end{align}
where (\ref{eqn:GradRyz}) follows since $\psi(1) = -\gamma$.

The third term (on the second line) in (\ref{eqn:MainIntegral}) becomes
\begin{align}
	\nonumber \lefteqn{c\int_0^\infty e^{-\lambda_W y} \log \left( 1 - e^{-(\lambda_W-\lambda_X )y} \right) dy} & \\
	\label{eqn:ThirdTerm}
	&= - \frac{\gamma + \psi \left( 1 + 
		\frac{\lambda_W}{\lambda_W - \lambda_X}\right)}{(\lambda_W - \lambda_X)/\lambda_X} .
\end{align} 
By a similar derivation, the fourth term (on the third line) in (\ref{eqn:MainIntegral}) becomes
\begin{align}
	\label{eqn:FourthTerm}
	\nonumber \lefteqn{c\int_0^\infty e^{-\lambda_X y} \log \left( 1 - e^{-(\lambda_W-\lambda_X )y} \right) dy} & \\
	&= 
	- \frac{\gamma + \psi \left( 1 + 
		\frac{\lambda_X}{\lambda_W - \lambda_X}\right)}{(\lambda_W - \lambda_X)/\lambda_W} .
\end{align}

Finally, we obtain
\begin{align}
	\nonumber \lefteqn{h(Y)} & \\
	\nonumber =&\:
	\frac{1 + \gamma + \psi\left( 1 + \frac{\lx}{\lw - \lx} \right) - \log \left(\frac{\lw\lx}{\lw - \lx}\right)}{(\lw - \lx)/\lw}\\
	\label{eqn:UnsimplifiedResult}
	& -\frac{\frac{\lx}{\lw}+\gamma + \psi\left( 1 + \frac{\lw}{\lw - \lx} \right) - \log \left(\frac{\lw\lx}{\lw - \lx}\right)}{(\lw - \lx)/\lx} .
\end{align}

This expression can be further simplified. For terms that appear in both numerators (for example the $\gamma$ term, but also the $\log$ term),
we have
\begin{equation}
	\frac{\gamma}{(\lw-\lx)/\lw} - \frac{\gamma}{(\lw-\lx)/\lx} = \gamma .
\end{equation}
Further, considering the leading term of the second numerator,
\begin{align}
	\nonumber \lefteqn{\frac{\frac{\lx}{\lw}}{(\lw - \lx)/\lx}} & \\
	&= \frac{\frac{\lx}{\lw} - 1 + 1}{(\lw - \lx)/\lx} \\
	&= \frac{\lx - \lw}{(\lw - \lx)\frac{\lw}{\lx}} + \frac{1}{(\lw - \lx)/\lx} \\
	&= - \frac{\lx}{\lw} + \frac{1}{(\lw - \lx)/\lx} .
\end{align}
From the above simplifications, and
noting that 
\begin{equation}
1 + \frac{\lx}{\lw-\lx} = \frac{\lw}{\lw-\lx} ,
\end{equation}
we can rewrite (\ref{eqn:UnsimplifiedResult}) as
\begin{align}
\nonumber \lefteqn{h(Y)} & \\ 
\nonumber =& \: 
	1 + \gamma + \frac{\lx}{\lw} + \log\left(\frac{\lw-\lx}{\lw\lx}\right) \\
	& +
	\frac{\lw \psi\left(\frac{\lw}{\lw - \lx} \right) - \lx \psi\left( 1 + \frac{\lw}{\lw - \lx} \right)}{\lw - \lx} .
\end{align}
A property of the digamma function is that \cite{weisstein-book}
\begin{equation}
	\label{eqn:DigammaProperty}
	\psi(1 + x) = \psi(x) + \frac{1}{x} .
\end{equation} 
Thus, we have
\begin{align}
	\nonumber
	\lefteqn{h(Y)} & \\
	\nonumber =& \: 1 + \gamma + \frac{\lx}{\lw} + \log\left(\frac{\lw-\lx}{\lw\lx}\right) \\
	& +
	\frac{\lw \psi\left(\frac{\lw}{\lw - \lx} \right) - \lx \psi\left( \frac{\lw}{\lw - \lx} \right) - \frac{\lx}{\lw}(\lw-\lx)}{\lw - \lx} \\
	&= 1 + \gamma + \log\left(\frac{\lw-\lx}{\lw\lx}\right) + \psi\left( \frac{\lw}{\lw - \lx} \right) ,
\end{align}
which is identical to (\ref{eqn:MainResult}).

\section{Examples}

\subsection{Illustration of the result}

\begin{figure}[t!]
\begin{center}
\includegraphics[width=3.25in]{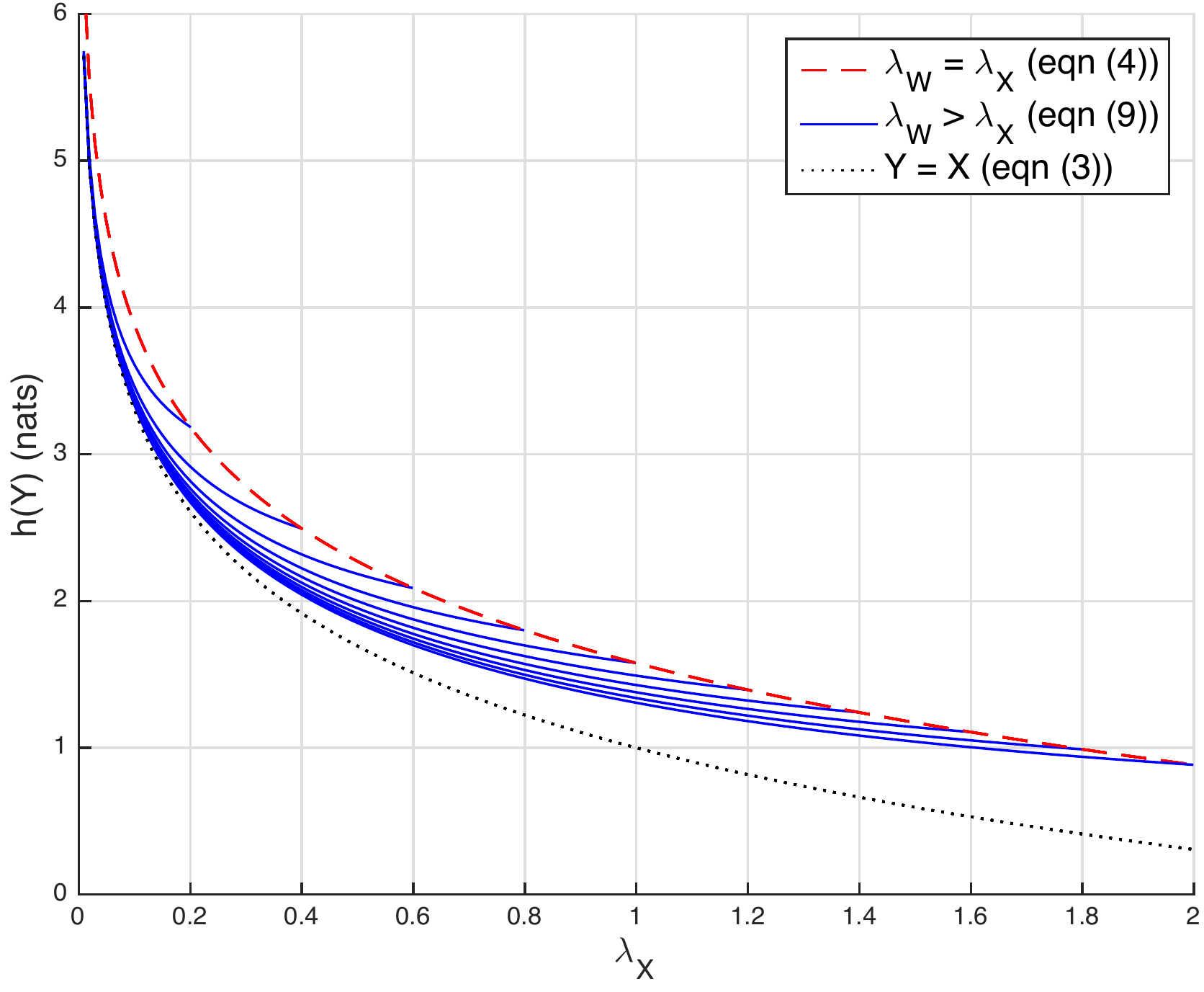}
\end{center}
\caption{\label{fig:EntropyExample} Plot illustrating the main result. The dashed line depicts
the entropy of $Y = W+X$, where $W$ and $X$ are IID exponential random variables ($\lambda_W = \lambda_X$), using (\ref{eqn:IdenticalEntropy}). The
solid lines depict the output of (\ref{eqn:MainResult}), where $\lambda_W > \lambda_X$, starting with $\lambda_W = 0.2$ in the
top line, and proceeding in increments of 0.2 to $\lambda_W = 2$ in the bottom line. The dotted
line depicts (\ref{eqn:SingleEntropy}), where $Y = X$ (i.e. a single exponential random variable with parameter $\lambda_X$).}
\end{figure}

In Figure \ref{fig:EntropyExample}, we plot the differential entropy $h(Y)$, given by (\ref{eqn:MainResult}), as a function of 
$\lambda_X$ and $\lambda_W$. This plot is given in comparison with
(\ref{eqn:SingleEntropy}) and (\ref{eqn:IdenticalEntropy}).

\begin{figure}[h!]
\begin{center}
\includegraphics[width=3.35in]{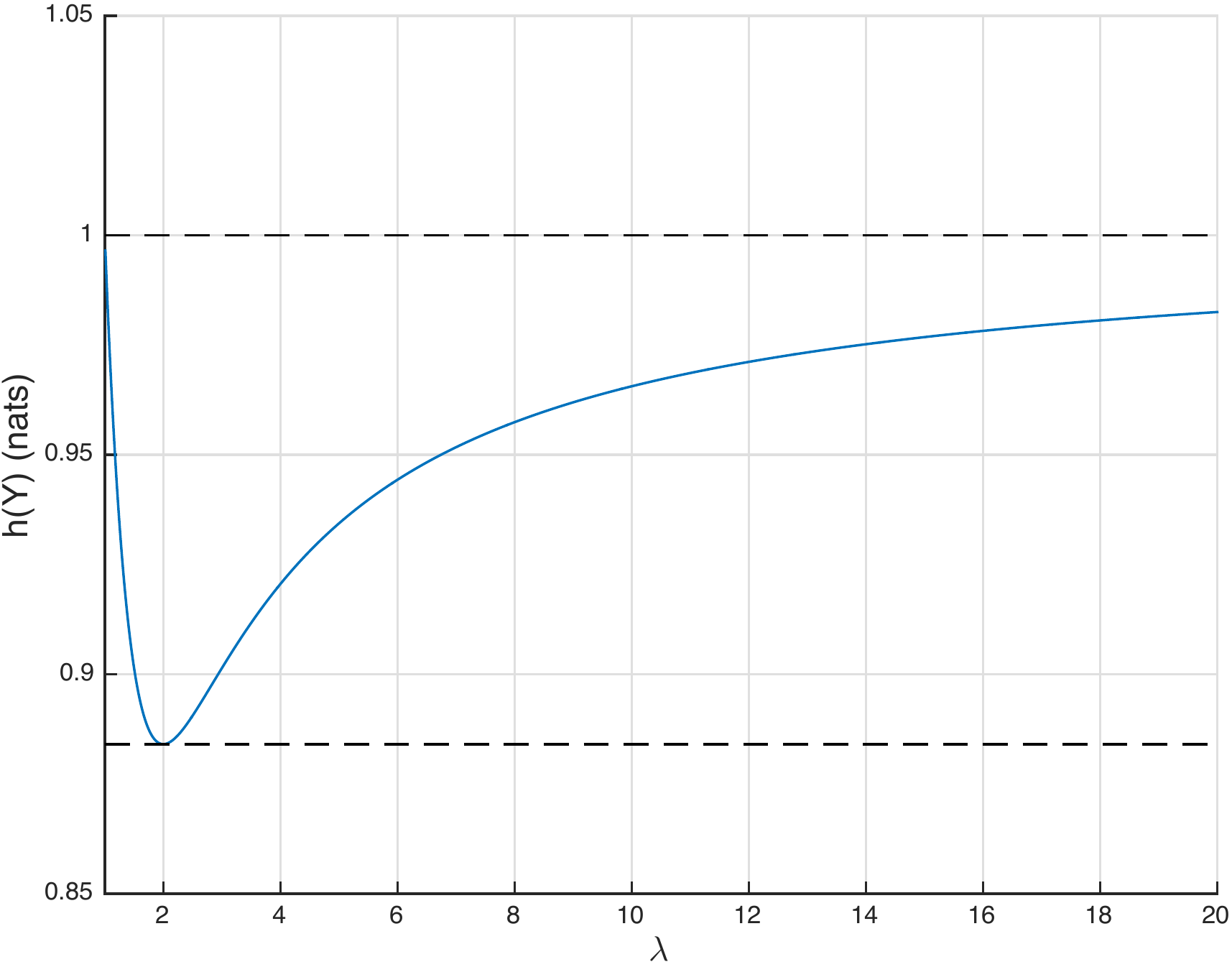}
\end{center}
\caption{\label{fig:EntropyExample2} 
Plot depicting constant expected value.
The blue line depicts (\ref{eqn:MainResult}) where $\lx = \min\{\lambda,\lambda/(\lambda-1)\}$ 
and $\lw = \max\{\lambda,\lambda/(\lambda-1)\}$, where $\lambda$ is given on the horizontal
axis.
The top dashed line depicts entropy (\ref{eqn:SingleEntropy}) for an exponentially-distributed random variable $Y$ with 
$\lambda = 1$.
The bottom dashed line depicts entropy (\ref{eqn:IdenticalEntropy}) for an Erlang-2-distributed random variable $Y$ with 
$\lambda = 2$. For all distributions, $E[Y] = 1$.}
\end{figure}

In Figure \ref{fig:EntropyExample2}, we plot the differential entropy $h(Y)$ 
given by (\ref{eqn:MainResult}), (\ref{eqn:SingleEntropy}), and (\ref{eqn:IdenticalEntropy}),
where the mean is constrained for each distribution such that $E[Y] = 1$. 
For our distribution in (\ref{eqn:NonIdenticalPDF}), this occurs when 
$\lx = \min\{\lambda,\lambda/(\lambda-1)\}$ 
and $\lw = \max\{\lambda,\lambda/(\lambda-1)\}$, for $\lambda \in (1,\infty)$.
The entropy in (\ref{eqn:SingleEntropy}) provides an upper bound on (\ref{eqn:MainResult}),
since the exponential distribution has maximum entropy for all distributions with the same mean;
the entropy in (\ref{eqn:MainResult}) approaches (\ref{eqn:SingleEntropy}) as $\lambda \rightarrow 1$
and $\lambda \rightarrow \infty$.

%
%

\subsection{Mutual information in additive exponential noise channels}

Consider the following covert timing channel over a network, along the lines of \cite{wagner2005}. Suppose 
a covert transmitter manipulates the time $X$ of a server request to send covert information;
in this example, we imagine that $X$ must have an exponential distribution with rate $\lx$, in order to avoid 
suspicious behaviour.
Further suppose the service time is $W$, which is an exponentially distributed random variable
with rate $\lw$.
Finally, suppose that the covert receiver can observe only the server's response time $Y$.
Thus, we have an additive exponential noise channel
with $Y = X + W$. 

Since $X$ and $W$ are both exponentially distributed,
then we can use the result in this letter to find a closed-form expression for the mutual information $I(X;Y)$, i.e. the highest rate of reliable communication, measured in nats per server request.
Let $\lx$ and $\lw$ be the rates of the exponentially-distributed $X$ and $W$,
respectively. If $Y = X+W$, and $\lx \neq \lw$, then $Y$ has the distribution given in (\ref{eqn:NonIdenticalPDF}). The mutual information is
\begin{align}
	I(X;Y) &= h(Y) - h(Y \given X)\\
	\label{eqn:MutualInfoExample1}
	&= h(Y) - h(W),
\end{align}
where (\ref{eqn:MutualInfoExample1}) follows since, given $X$, we can obtain $W = Y-X$ (this is a well-known property of additive noise channels).
Assuming $\lw > \lx$, we have
\begin{align}
	\nonumber\lefteqn{I(X;Y)}&\\ 
	\nonumber&= 1 + \gamma + \log\left(\frac{\lw-\lx}{\lw\lx}\right) + \psi\left( \frac{\lw}{\lw - \lx} \right)\\
	& \:\:- 1 + \log \lw\\
	\label{eqn:MutualInfoExample2}
	&= \gamma + \log\left(\frac{\lw-\lx}{\lx}\right) + \psi\left( \frac{\lw}{\lw - \lx} \right) .
\end{align}

Verd\'u \cite{verdu96} gives the capacity of the additive exponential noise
channel, and finds that the mean-constrained capacity-achieving distribution is a mixture of a point mass with an exponential distribution. 

\subsection{Conditional entropy in continuous-time Markov systems}

Most signal transduction systems can be modelled as continuous-time Markov processes \cite{eckford16}. For example,
Channelrhodopsin-2 (ChR2) is a light-sensitive protein with
numerous biological and bioengineering applications.
ChR2 operates by opening an ion channel in response to absorbing a photon.
In one common model of ChR2 \cite{nagel03}, 
the protein has three states: open, degraded, and closed.
The ion channel is only open during the open state, and the receptor must
pass through the degraded and closed states before opening again.
However, the transition rate from closed to open is dependent on the light intensity.

If $Y$ is the time between channel openings, $X$ is the duration of the 
degraded state, and $W$ is the duration of the closed state, we have $Y = X+W$;
moreover, since the system is Markov, $X$ and $W$ have the exponential distribution:
$X$ has rate $\lx$, and $W$ has rate $\lw^{(L)}$, where $L$ is the light intensity.
In experiments, the light is often either on or off; thus, we will use $\lw^{(\mathrm{on})}$
and $\lw^{(\mathrm{off})}$ to represent the corresponding rates. 

We can use the result in this letter to obtain the conditional entropy of $Y$ given $L$.
Typically, $\lw^{(\mathrm{off})} < \lx < \lw^{(\mathrm{on})}$, so we have
\begin{align}
	\nonumber\lefteqn{h(Y \given L)}&\\ &= p(\mathrm{off}) H(Y \given L = \mathrm{off}) +
		p(\mathrm{on}) H(Y \given L = \mathrm{on}) \\
	\nonumber &= 1 + \gamma \\
	\nonumber &\:\:\: + p(\mathrm{off}) \left[ \log\left(\frac{\lx-\lw^{(\mathrm{off})}}{\lw^{(\mathrm{off})}\lx}\right) + \psi\left( \frac{\lx}{\lx-\lw^{(\mathrm{off})}} \right) \right]\\
	&\:\:\:+p(\mathrm{on}) \left[ \log\left(\frac{\lw^{(\mathrm{on})}-\lx}{\lw^{(\mathrm{on})}\lx}\right) + \psi\left( \frac{\lw^{(\mathrm{on})}}{\lw^{(\mathrm{on})} - \lx} \right) \right] .
\end{align}
This intermediate result can be used to calculate the mutual information $I(Y;L)$.

\subsection{Limit as $\lx \rightarrow \lw$}

Finally, we can show that the entropy of the Erlang-2 distribution (\ref{eqn:IdenticalEntropy}) emerges from our main result, using standard limits
and properties of the digamma function.

We will want the difference between the two exponential rates to vanish, so we change the notation
slightly: let $\lambda_\Delta = \lambda_W - \lambda_X$, and further let $\lambda_X = \lambda$ and
$\lambda_W = \lambda + \lambda_\Delta$.
Then, using (\ref{eqn:DigammaProperty}),
\begin{align}
\nonumber\lefteqn{h(Y)}&\\ &= 1 + \gamma + \log\left(\frac{\lambda_\Delta}{(\lambda + \lambda_\Delta) \lambda} \right)
+ \psi \left( \frac{\lambda + \lambda_\Delta}{\lambda_\Delta} \right) \\
&= 1 + \gamma - \log (\lambda + \lambda_\Delta) - \log \left( \frac{\lambda}{\lambda_\Delta} \right)
+ \psi \left( 1 + \frac{\lambda}{\lambda_\Delta} \right) \\
&= 1 + \gamma - \log( \lambda + \lambda_\Delta) - \log \left( \frac{\lambda}{\lambda_\Delta} \right)
+ \frac{\lambda_\Delta}{\lambda} + \psi \left( \frac{\lambda}{\lambda_\Delta} \right)
\end{align}
and
\begin{align}
	\nonumber \lefteqn{\lim_{\lambda_\Delta \rightarrow 0} h(Y)}&  \\
	\label{eqn:LambdaDeltaLimit}
	&= 1 + \gamma - \log \lambda + \lim_{\lambda_\Delta \rightarrow 0} \left[ \psi \left( \frac{\lambda}{\lambda_\Delta} \right) - \log \left( \frac{\lambda}{\lambda_\Delta} \right)\right] \\
	\label{eqn:LimitExample1}
	&= 1 + \gamma - \log \lambda ,
\end{align}
where the last line follows from limits of the digamma function.

From (\ref{eqn:DigammaProperty}), and using the property that $\psi(1) = - \gamma$,
\begin{align}
	2 - \psi(2) &= 2 - \psi(1+1)\\
	&= 2 - 1 - \psi(1) \\
	&= 1 + \gamma
\end{align}
and (\ref{eqn:LimitExample1}) becomes
\begin{equation}
	\lim_{\lambda_\Delta \rightarrow 0} h(Y) = 2 - \psi(2) - \log \lambda
\end{equation}
which is identical to (\ref{eqn:IdenticalEntropy}).

Note that this limit is depicted in Figure \ref{fig:EntropyExample2} at $\lambda = 2$,
where the curve touches the bottom line.

%

\bibliographystyle{ieeetr} 

\end{document}